\documentclass[useAMS,usenatbib]{mn2e}

\usepackage{natbib}
\usepackage{graphicx}
\usepackage{rotating}
\usepackage{times}
\usepackage{amsmath}

\title[Magellanic Stream modeling]{Lessons from the Magellanic System and its modeling} 

\author[Jianling Wang et al.]{
  Jianling Wang$^{1}$\thanks{E-mail:wjianl@bao.ac.cn},
  Francois Hammer$^{2}$\thanks{E-mail:francois.hammer@obspm.fr},
  and Yanbin Yang$^{2}$\\
{$^1$ CAS Key Laboratory of Optical Astronomy, National Astronomical Observatories, Beijing 100101, China,} \\
{$^2$ GEPI, Observatoire de Paris, CNRS, Place Jules Janssen 92195, Meudon, France.}\\
}

\begin{document} 

\date{Received ; accepted}

\maketitle

\begin{abstract}

The prominent Magellanic Stream that dominates the HI sky provides a tantalizing number of observations that potentially constrains the Magellanic Clouds and the Milky Way outskirts. Here we show that the 'ram-pressure plus collision' model naturally explain these properties, and is able to predict some of the most recent observations made after the model  was made. These include the complexity of the stellar populations in the  Magellanic Bridge, for which kinematics, ages, and distances are well measured, and the Northern Tidal Arm, for which the model predicts its formation from the Milky Way tidal forces. It appears that this over-constrained model provides a good path to investigate the Stream properties. 
This contrasts with tidal models that reproduce only half of the Stream's main properties, in particular a tidal tail cannot reproduce the observed inter-twisted filaments, and its gas content is not sufficiently massive to provide the large amount of HI and HII gas associated to the Stream. Despite the efforts made to reproduce the large  amounts of gas brought by the Clouds, it seems that no viable solution for the tidal model could be foreseen. Since the 'ram-pressure plus collision' model has not succeeded for a Large Magellanic Cloud mass above 2 $\times10^{10}$ $M_{\odot}$, we conjecture that a low mass is required to form the Stream.

\end{abstract}


\section{Introduction}

Together, the Magellanic Stream (MS) and  Leading Arm (LA) subtend an angle of 230 degree, making it the second most prominent neutral hydrogen structure dominating the sky, after the Milky Way.  The MS has been identified to be anchored to the Magellanic Clouds in 1974 by \citet{Mathewson1974}, though the nature of its formation was considered still unknown in 2012 \citep{Mathewson2012}. HST proper motion measurements of the Magellanic Clouds by \citep{Kallivayalil2006,Piatek2008,Kallivayalil2013}  indicate that the Clouds are presently moving at high velocities, and are consistent with a first passage about the Galaxy \citep{Besla2007}. In such a frame, modeling of the MS has followed two very different and mutually exclusive schemes, since either it can be a tidal tail \citep[and references therein]{Besla2012}, or it can be made by one or two ram-pressure tails \citep{Mastropietro2010,Hammer2015}.\\

Today the most influential tidal model of the MS is that of \citet{Besla2012}, for which the MS is a gigantic tidal tail extracted from the Small Magellanic Cloud (SMC) by the tidal effect of the Large Magellanic Cloud (LMC) during a close interaction 1.2 to 2 Gyr ago. The SMC is further assumed to be a long-lived satellite of the LMC, which imposes a very large mass for the latter, in excess of $10^{11}$ $M_{\odot}$  \citep{Kallivayalil2013}. The strengths of the \citet{Besla2012} model are their predictions of the MS length, of the 6 D space-velocity phase of the Clouds, and that it predicts at least one arm of the LA. 

The major limitations of the \citet{Besla2012} tidal model include (i) that only a small fraction, few percent, of the MS gas is reproduced, and no ionized gas while it is the prominent component of the MS \citep{Fox2014}, (ii) the stream is made of two filaments, with kinematic and chemical analyses indicating that gas from both the LMC and SMC is present, and (iii) the absence of stars in the Stream observed so far\footnote{\citet{Besla2013} argued that the absence of stars could be due to  their expected very low surface brightness, especially if only few stars were extracted from a very gas-rich  SMC. \citet{Zaritsky2020} recently claimed a positive detection of stars near the MS tip end, which is at odd with the \citet{Besla2013}'s argument, since the MS furthest part is likely at its tip end. However, this detection of MS stars needs confirmation, because they are offset from the HI gas stream and their distance are much lower than model's expectations.}. Each of these limitations appear to be a showstopper for the tidal model of the MS. 
 
 The \citet{Besla2012} model is also limited because it neglects the presence of a coronal phase in the Milky Way (MW) halo, while the presence of a multi-phase circumgalactic medium (CGM) has been evidenced by many means (X-rays: \citealt{Bregman2018,Faerman2017, Miller2013}; QSO absorption lines:  \citealt{Zheng2015,Zheng2019,Fox2014,Richter2017}; LMC shrunk gas disk: \citealt{Nidever2014}; high velocity cloud dissociation: \citealt{Kalberla2006}). More recent realizations of the tidal model have introduced the MW coronal phase together with that of a putative LMC corona \citep{Lucchini2020,Lucchini2021}. \\


Compared to tidal models, the 'ram-pressure plus collision' model \citep{Hammer2015,Wang2019} appears much more advanced since it reproduces in details many known properties of the MS, as well as some qualitative. In this model the LA is produced by the leading passages of gas-rich dwarfs assumed to be progenitors of present-day MW dwarfs (see also \citealt{Tepper-Garcia2019}). This hypothesis has been also adopted by the most recent tidal model of the MS \citep{Lucchini2021}, and is further supported by the  MW dwarf high energies and angular momenta, suggesting also a recent passage for them \citep{Hammer2021}.
By construction, the 'ram-pressure plus collision' model reproduces the dual filamentary structures, which are two ram-pressure tails attached to each Clouds, the whole HI MS shape and gas mass and its radial velocity \citep{Hammer2015}. The recent collision between the Clouds explains the Magellanic Bridge, leading to Cloud proper motions consistent with the observed values. It also reproduces the velocity field of the LMC and the gigantic, 30 kpc-long,  structure along the line of sight of the SMC young stars \citep{Wang2019,Ripepi2017}, which both are due to the recent, 300 Myr old, collision between the gas-rich Clouds. Moreover, the MS ionized gas deposited during the MC motions in the MW halo is coming from the HI gas of the Clouds, which has been extracted by ram-pressure, and then ionized  by the hot corona of the MW \citep{Wang2019}.\\

However and quite surprisingly, the 'ram-pressure plus collision' model does not assume a dark matter component for both Clouds, while it naturally reproduces the MS in details without fine-tuning. It even constrains the LMC mass to be smaller than 2 $\times 10^{10}$ $M_{\odot}$, because otherwise it would not let sufficient expelled gas to make the mass of the HI and especially that of the HII MS. This may come at odd with LMC mass estimates based on the interaction between the MW and the LMC \citep{Erkal2019,Erkal2021,Conroy2021,Vasiliev2021}. \\

In this paper we aim at testing the 'ram-pressure plus collision' model by considering the most recent and new constraints on the Magellanic System. The tests include the morphological-age-kinematics distribution of stars in the Magellanic Bridge (MB), and the morphological and kinematic behavior of the faint Northern Tidal Arm (NTA), which stretch an angle of 12.5 degree. In principle, the tantalizing number of observations that are reproduced makes this model over-constrained and hence, able to be predictive. 
The paper is organized as it follows. In section 2, we describe the observational properties of two populations identified in the MB region, the 'ram-pressure plus collision' model from \citet{Wang2019}, comparison between MB stellar populations and predictions from the simulations ,as well as an interpretation of  their origin. In section 3, we compare the Gaia results for  the NTA with our simulation model predictions. In section 4, we discuss the advantages and inconveniences of the different Magellanic System modeling, and then conclude in section 5.

\section{The two stellar populations in the Magellanic Bridge}

\subsection{Additional constraints from the Bridge}
In the MB region, the neutral gas bridge has been well identified by HI
observations \citep{Nidever2010}, which well elucidate the mutual interaction
between MCs.  A population of young stars in the Bridge has been also
identified by various observations \citep{Irwin1985,Demers1998}, and their
strong correlation with the MB neutral gas  indicates that they likely result
from in-situ star-formation during the recent MC interaction
\citep{Skowron2014,Belokurov2017}.

Besides young stars in the Bridge region,
older age RGB stars have been also discovered by
\citet{Bagheri2013} and \citet{Noel2013}, and \citet{Skowron2014} identified the
presence of red clump stars as well. These older age stars are either
distributed with a large scatter in the Bridge region or in front of the gas
Bridge region with respect to the motion direction (see also
\citealt{Belokurov2017}).

Recent observations of the MB have confirmed stellar populations with ages ranging from Young Main Sequence stars and old ancient RR Lyrae
\citep{Belokurov2017}. Based on star formation and orbital past histories of the Clouds, the MB has been formed $\sim 200-300$ Myr ago \citep{Casetti-Dinescu2012,Hammer2015}. Thanks to the $Gaia$
precise proper motion data, stellar tangential motions have been measured in
the MB region, revealing that stars in the MB are leaving SMC towards the LMC  \citep{Belokurov2017,Luri2020,Omkumar2020,Zivick2019,Schmidt2020}.
It suggests that they have been stripped from the SMC due to the  LMC tidal force. By using the red clump (RC) as standard candle, \citet{Nidever2013}
found that there are two stellar populations with different brightnesses within the MB, which has been
recently confirmed by \citet{Omkumar2020}. With $Gaia$ DR2 data, \citet{Omkumar2020} found that the bright star population at the SMC distance
has a larger tangential velocity than the fainter, more distant population. This observation may provide a useful constraint for models to simulate the formation of Magellanic System.


\subsection{Comparison between simulated data and observations}

\subsubsection{The ram-pressure plus collision model}

\citet{Wang2019} adapted a ''ram-pressure plus collision'' model to
reproduce the MS and MCs formation using the state-of-the-art software GIZMO
\citep{Hopkins2015}. In this model the MS is formed by an interaction between MCs
and ram-pressure exerted by the Milky Way hot corona. It leads to dual HI
streams behind each MCs, and explains the kinematics and the mass of the huge
amount of ionized gas kinematically coupled with the MS. Besides this, the
strong interaction between MCs tidally reshapes SMC to a very elongated
structure. 

In the model of \citet{Wang2019}, the stellar distribution of SMC consists
of a disk and a spheroid components. The stellar disk is an exponential
distribution with scalelength 1.5 kpc, and the spheroid component follows the
profile of \citet{Dehnen1993} with $\gamma=0$, which has a core in the center
with half mass radius 5.8 kpc.  The stellar disk represents the young star
population observed in the SMC, while the extended spheroid component is motivated by the
spheroidal distribution of ancient RR Lyrae stars observed around the SMC
\citep{Ripepi2017}. This model of the SMC have been shown to reproduce  well the very extended
cylindrical structure with line-of-sight distance around 30 kpc for young
stars, as well as the nearly spheroidal distribution of RR Lyrae
stars \citep{Ripepi2017,Scowcroft2016,Wang2019}. \citet{Wang2019} discussed  all details for the initial conditions and
modeling.

To compare the 3D morphology of MCs with RR Lyrae data and Classic Cepheid
observed in \citet{Ripepi2017}, \citet{Wang2019} have randomly selected
particles to match the numbers to the observed stars, and to match the observed
region. This renders it difficult to distinguish the faint features around MCs,
especially in the Magellanic Bridge region as shown in Fig. 7 of
\citet{Wang2019}.

Here, to have a view of the whole particle distribution in 3D space, 
we now show in Fig.\ref{fig:LBD} the simulated MCs particles distribution with all of particles used in our simulation model of \citet{Wang2019}. 
The green color
dots indicate the LMC particles, while the cyan color dots represent the SMC.
In this figure, particles distribute over a much larger region than in
Fig.7 of \citet{Wang2019}, which illustrates well that the MB is connecting both MCs.

\subsubsection{Two stellar populations at different distances in the Bridge}

\begin{figure*}
\center
\includegraphics[scale=0.42]{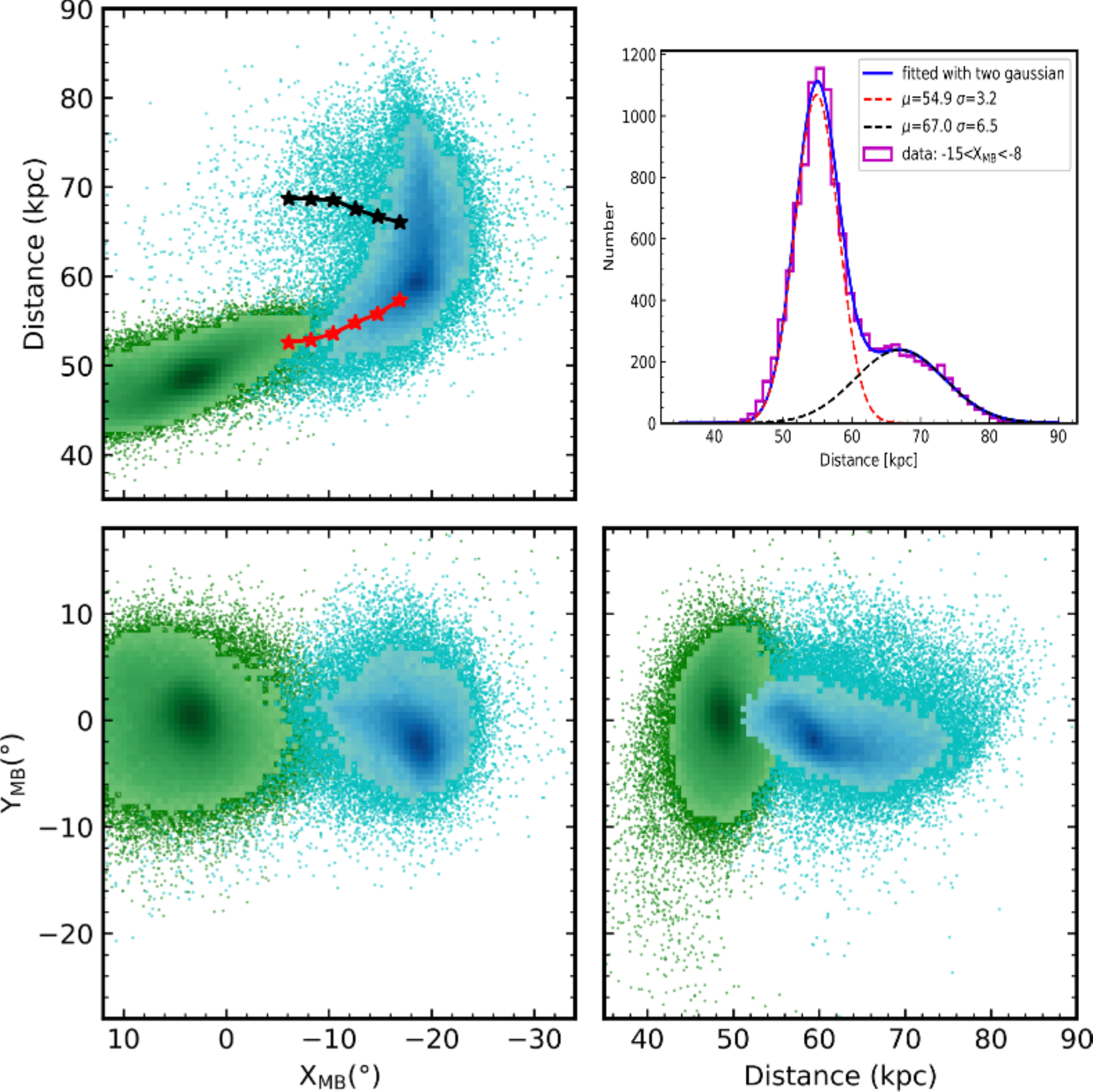}
\caption{Simulated particles distributions of LMC and SMC in Magellanic Bridge
coordinates \citep{Belokurov2017} (bottom-left), or Magellanic Bridge
longitude/latitude versus distance (bottom-right and top-left, respectively).
Greens (cyan) points are particles from LMC (SMC). In the top-left panel, the
red and black big stars indicates the mean distance values along Magellanic
Bridge longitude for the background and foreground population, which are
separated at 60 kpc. The top-right panel indicates the distance distribution 
for stars within the Bridge region, to which are overplotted two Gaussian functions 
fitted to this distribution and the fitted results are labeled on this panel. } 
\label{fig:LBD}
\end{figure*}

\begin{figure}
\center
\hspace*{-1cm}
\includegraphics[scale=0.68]{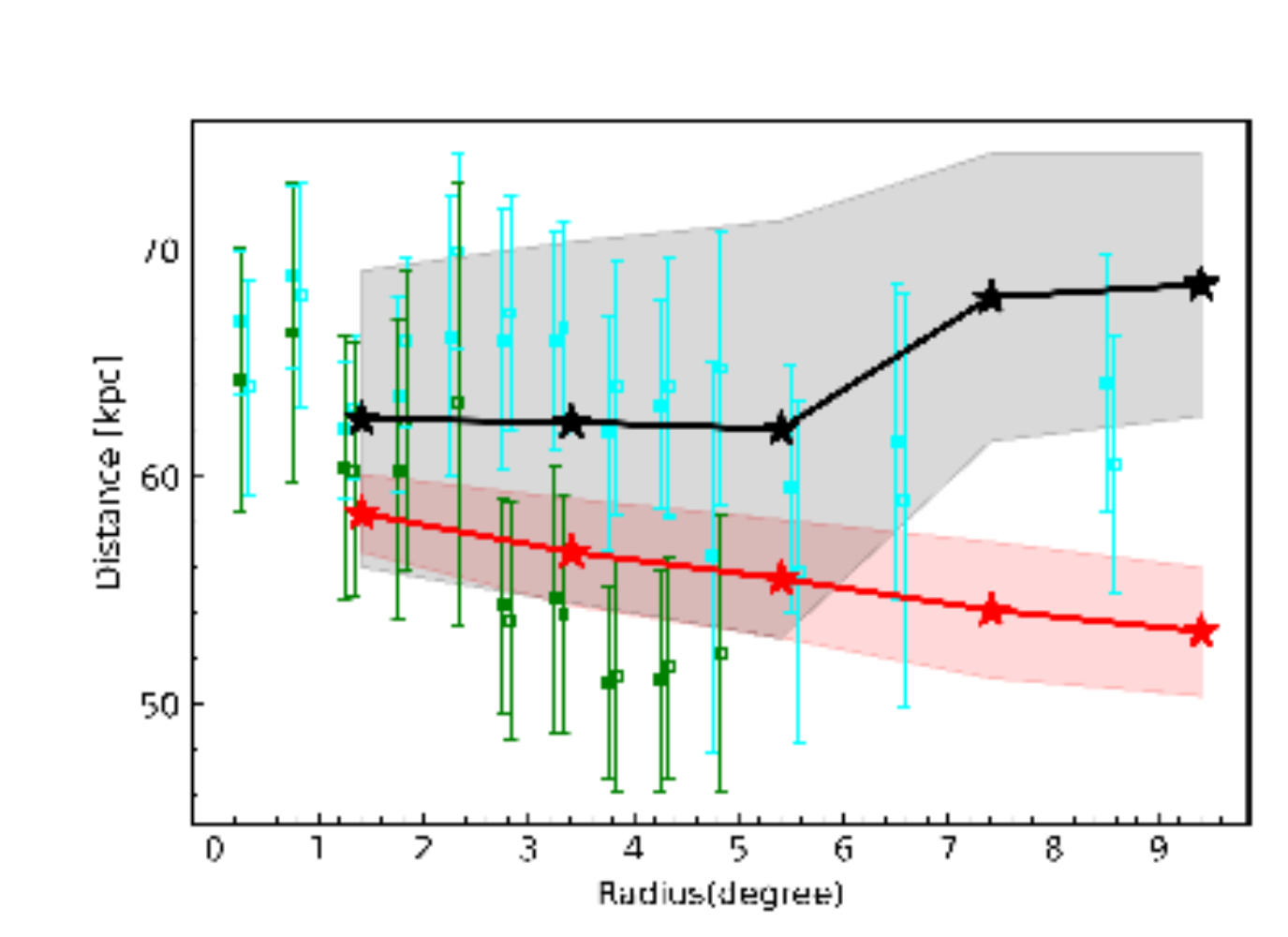}
\caption{Comparing relations of distances varied with radius to the SMC between
observation and simulations. Green and cyan color indicate bright and faint 
population of RC from \citet{Omkumar2020} for North East (solid square) and
South East region (open square). Simulation data from Model-52 
are shown by red and red stars, and the gray and 
pink region shown the 1$\sigma$ distance scatter.} 
\label{fig:ObsBins}
\end{figure}

\begin{figure*}
\center
\includegraphics[scale=0.40]{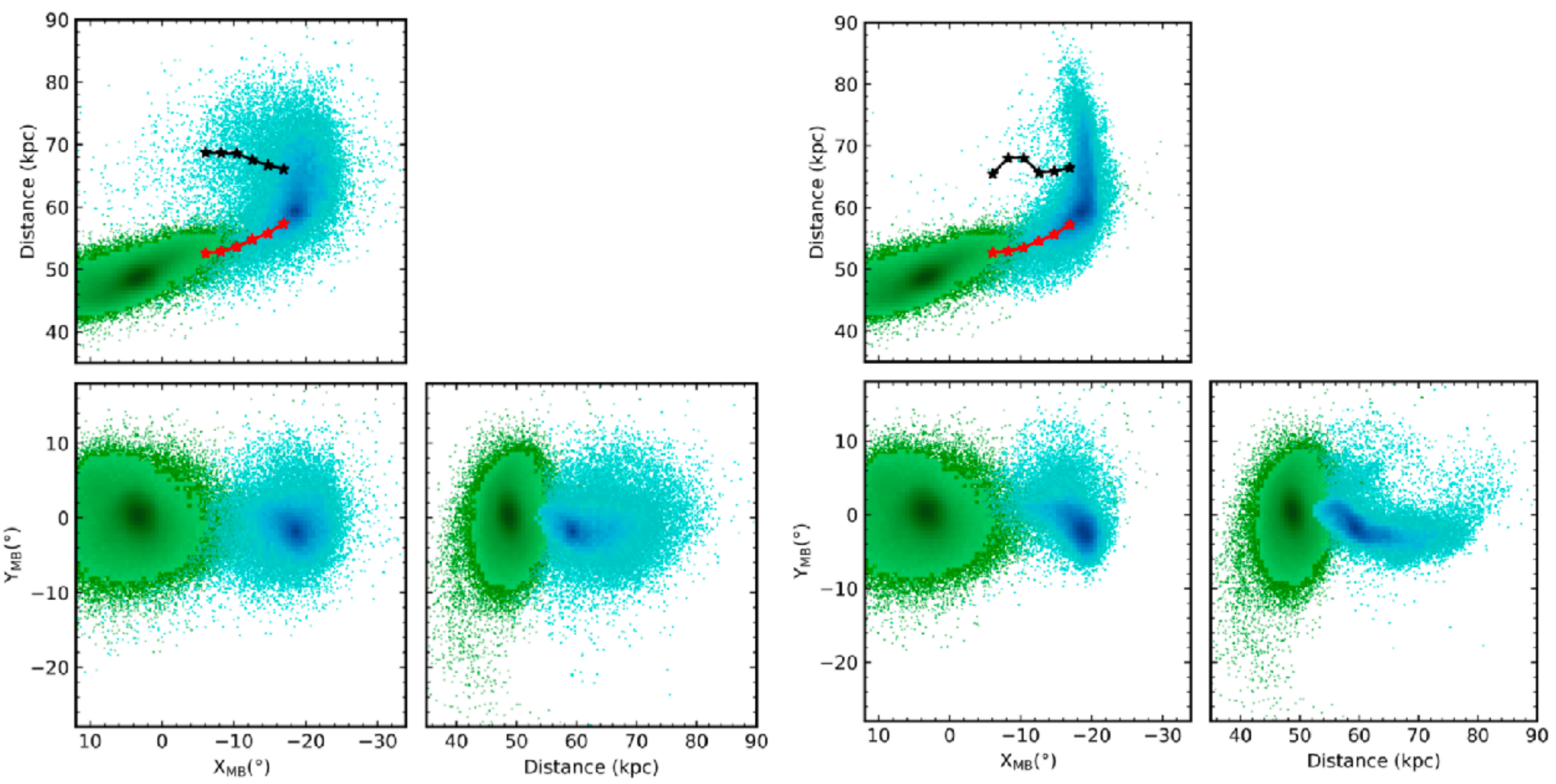}
\caption{Particles distributions in the MB longitude, latitude, and distance
space. The left panels show spheroid component of SMC and the right panels present
the disk component of SMC. The black and red stars indicate the mean distance along 
with the MB longitude for stars separated at 60 kpc.} 
\label{fig:BD}
\end{figure*}

Fig. \ref{fig:LBD} presents the whole star particle distribution in Magellanic
Bridge coordinates \citep{Belokurov2017} and also identifies their distances.
In Fig. \ref{fig:LBD} all particles are shown and their colors identify whether they
belong to the LMC (green) or to the SMC (cyan). In the X$_{\rm MB}$ versus
distance panel (top left panel), there is a tidal bridge (red dots) connecting
LMC to SMC with distance varying from $\sim$ 65 kpc (SMC) to $\sim$ 50 kpc
(LMC), which forms the stellar Magellanic Bridge (MB). Behind the MB,
there are particles of SMC distributed from 60 kpc to 80 kpc, which 
originated from the SMC after tidal interaction with LMC. The MB and background
populations of SMC are clearly separated in distance. 

In the top right panel of Fig. \ref{fig:LBD} the distance distribution of
particles within $-15{\degr} \la {\rm X_{MB}} \la -8{\degr}$ are shown with
a magenta line. There are two populations with different distances distribution.
Two Gaussian functions (red-dashed and black-dashed lines) are used to fit
these two populations, and the results for the mean and standard deviation of
distance are labeled on the top right of this panel. The two populations are
clearly located at two different distances, one at $\sim 55$ kpc that delineates the
mean value of the MB, the other one having a broader distance
distribution with a mean value of $\sim 67$ kpc. The two populations are well
separated at $\sim 60$ kpc.

\citet{Omkumar2020} used RC stars to trace distance and found that the stellar
population in the East part of SMC have two different populations at different
distances. To better compare with observation data, we fitted two Gaussian
functions with distance distribution for different MB longitude bins.  The
fitted results are shown in the Fig. \ref{fig:hist} of Appendix. All of the
distance distributions at different bins are well fitted with two Gaussian
functions. We note that the foreground population between 50 and 60 kpc is the
dominant component for all the distribution bins, while it is less prominent
in the observation \citep[see figure 4 of][]{Omkumar2020}. This indicates that
the interaction between LMC and SMC is so strong in the simulation that too many particles are
tidally stripped from the SMC. A fine tuning model parameters is needed to reduce
this component to match with observation. For example, this can be done by either increasing the pericenter to
decrease the tidal force, or by decreasing the size or the mass of the SMC disk component,
or to change the SMC inclination angle relatively to the orbital plan in order to decrease
resonance. 
 
Fig.\ref{fig:ObsBins} compares the distance variation at different SMC radii for the two components, 
allowing to compare observation data and simulation results. The
observed data are shown in green (bright RC population) and cyan (faint RC 
population), with solid squares for the North East and open squares for the South East.
The two Gaussian fitted results to the simulation data are shown with red
(foreground population) and black (background population) colors. The
simulated foreground population follow the trends that its distance decreases
with increasing SMC radius, which matches with the observed bright RC (green square)
population. The background population in the simulation has a nearly flat distance distribution
at radii less than $\la 7\degr$, and the distance increases at radii $\ga 7\degr$, which
is roughly similar to the faint RC population.

In our model, the progenitor of SMC consists of a disk and a spheroid component
\citep{Wang2019}.  In Fig. \ref{fig:BD} the spheroid (left panels) and disk
(right panels) components of SMC are separated and shown in the MB coordinates
or longitude/latitude versus distance. The background population are mainly
coming from the spheroidal component, while the foreground component includes contributions from
both the spheroid and the disk component.

\subsubsection{Kinematic features for the two populations}

\begin{figure*}
\includegraphics[scale=0.65]{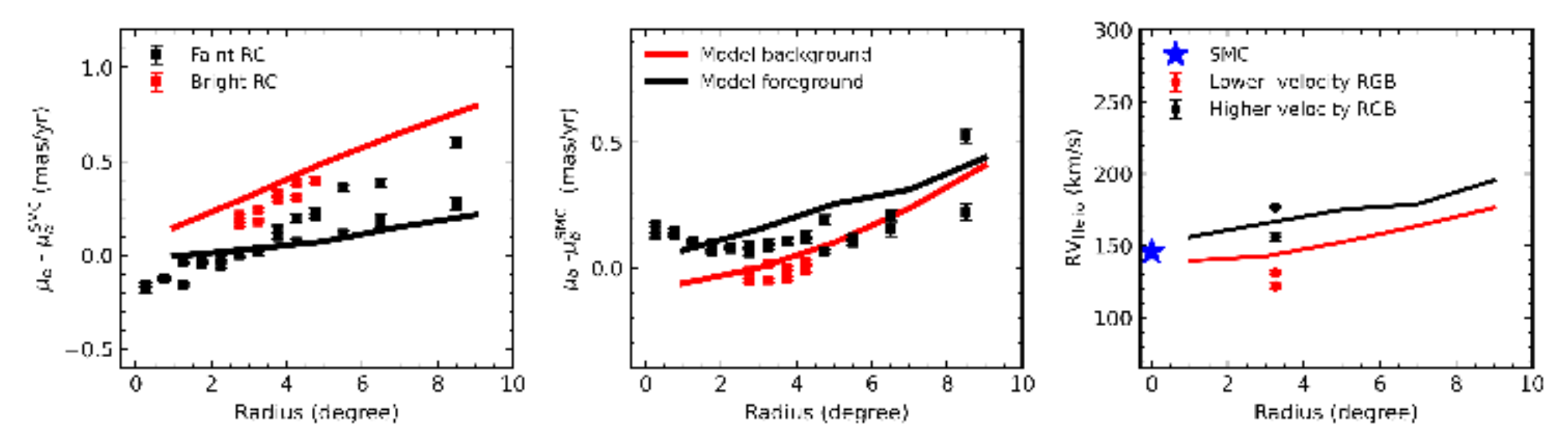}
\caption{
The relative variation of proper motions with respect to that of SMC as
function of radius to SMC are shown in left and middle panels, and the right
panel shows the radial velocity variation as function of radii to SMC. The
observed proper motions and radial velocity of SMC are from \citet{Zivick2018}.
The proper motion of faint (black square) and bright (red square) RC in the
left and middle panel are from \citet{Omkumar2020}. The radial velocity of the
lower (red circles) and higher (black circles) velocity RGB stars are from
\citet{James2021}, which are corresponds to foreground components and main body of SMC.
}
\label{fig:Kine} 
\end{figure*}

Using $Gaia$ DR2, \citet{Omkumar2020} studied the proper motions of these
two populations, and found two populations with different kinematics. They
identified the proper motions of the two populations in both directions
(${\mu}_{\alpha}$, ${\mu}_{\delta}$) as well as their variations against the SMC radius.

Left and middle panels of Fig.\ref{fig:Kine} show the relative proper motions variation of the two populations with respected to SMC 
as function of radius to SMC. The background (faint) and foreground (bright) RC are shown with black and red squares, while the modelled 
background and foreground are shown with black and red solid lines. The background (faint) RC component 
has been argued to belong to the main body of SMC \citep{Omkumar2020,James2021}, but this has to be clarified. This is because there are systematic differences ($\sim 0.16$ mas/yr) of ${\mu}_{\alpha}$ and ${\mu}_{\delta}$ near the central regions (see black squares at zero radius in left and middle panels of Fig.\ref{fig:Kine}. 
The modeled proper motions reproduce qualitatively the observations by many aspects, including the variations of proper motions
in both directions with the SMC radius. Both observations and simulations show that the foreground population has larger
$\mu_{\alpha}$ than the background population (left panel), and that this relation
is inverted in the $\mu_{\delta}$ (middle panel). The right panel of
Fig.\ref{fig:Kine} shows the radial velocity as function of radius for the
model. Both the background and foreground population shows increasing
velocity with increasing radius, namely it increases toward to LMC. The background population
have higher radial velocity than foreground population.
With $Gaia$ EDR3 and collected radial velocity data from literature, \citet{James2021} identified two radial
velocity components in the eastern region of SMC for the RGB stars with
velocity difference by $\sim 35-45$ km/s. The higher (lower) radial velocity
component has proper motions consistent with the background (foreground) RC
population. This suggests that the lower velocity component belongs to the same substructure than
the foreground population, and the higher velocity component corresponds to the
main body of SMC as well as the background component. In the right panel of
Fig.\ref{fig:Kine} the two radial velocity components of RGB from
\citet{James2021} are shown by black and red circles. Even though the difference 
of radial velocity between background and foreground in our model is smaller than that observed, 
our model correctly predicts that the foreground component has a lower velocity than that in thebackground.

\begin{figure*}
\includegraphics[scale=0.55]{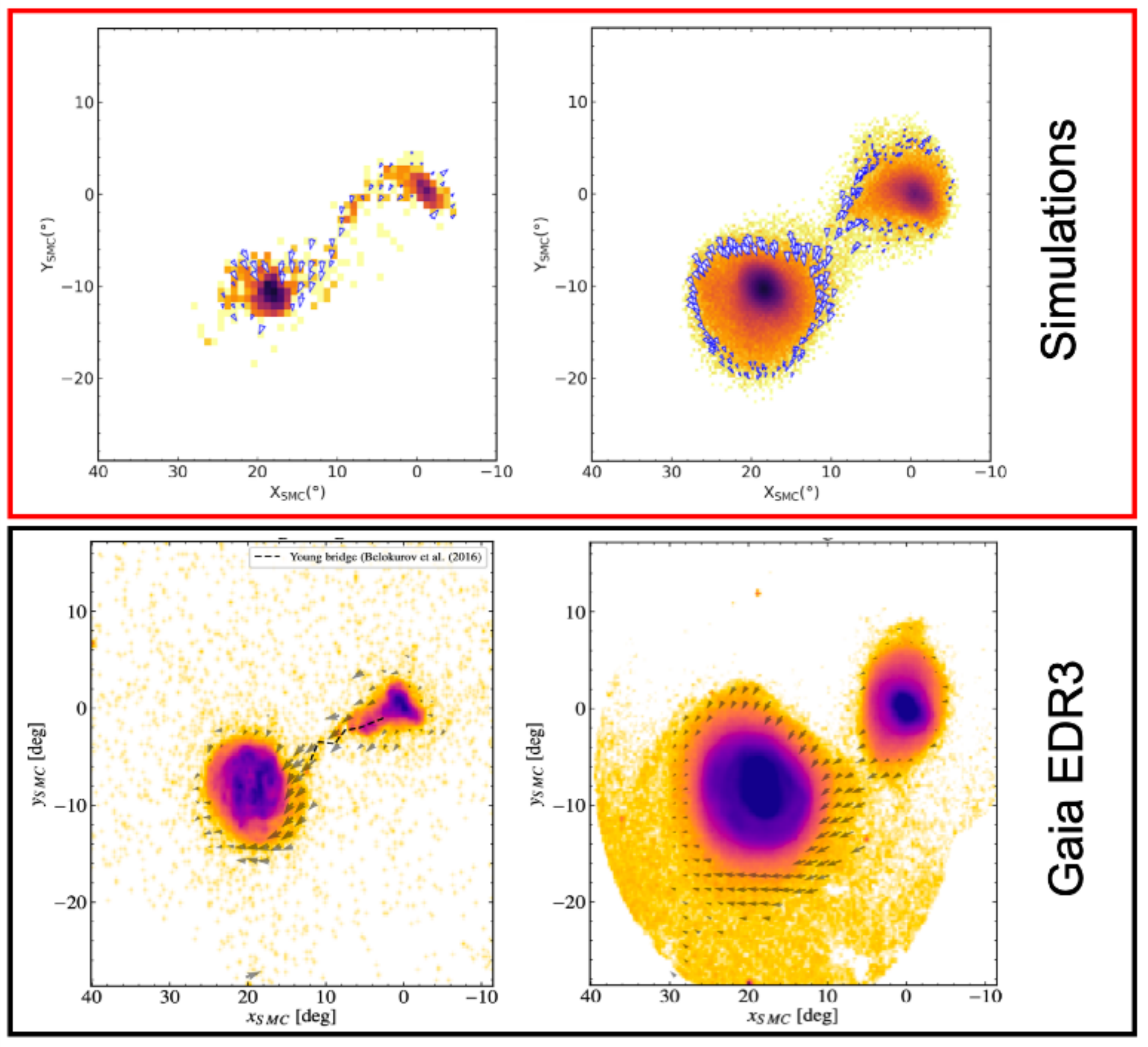}
\caption{Comparing the young and old population bridge between simulation model
and $Gaia$ EDR3 data. Vector field are overlapped on the density maps for young
stars (left panels) and old stars (right panels). The coordinates has been
centered on SMC for comparison with new results from $Gaia$ EDR3
\citep{Luri2020}. There is clearly young star bridge connecting LMC and SMC,
and moving toward to the LMC, which are consistent with the $Gaia$ EDR3 results
(bottom panel) for both density map and velocity vector \citep{Luri2020}. There
is also old stellar bridge for the observation and simulation results. Since
there are no quantitative length scale of proper motion vector available for the observation data from \citet{Luri2020}, 
the length scale of arrows are not matched between observation and model. }
\label{fig:PMsep} \end{figure*}

\citet{Luri2020} have confirmed the above observational results with $Gaia$ EDR3 data, showing that
there are both young and old stellar bridge having both their proper motions towards the
LMC.  To have a comparison, we selected young population as required stars with
age smaller than 150 Myr, and old stars with age larger than 2 Gyr which are
corresponding to the intermediate-old red clump population used in
\citet{Luri2020}. The top left-panel of Fig.\ref{fig:PMsep} shows the young
stars distribution on the sky, and the blue arrows indicate the proper motion
vector with respect to SMC. In the top right-panel of Fig.\ref{fig:PMsep} the old population shows a bridge connection between the LMC and the SMC, which has proper motions
similar to that of young stars moving from the two Clouds. For comparison, data from
$Gaia$ EDR3 are shown in the bottom panels of Fig.\ref{fig:PMsep} from
\citet{Luri2020}. Both young and old population in the Bridge region share
similar motion properties for both observations and model, in particular the Bridge star motions 
towards the LMC. Only qualitative comparison is feasible because there is no available length scale for the proper motion vectors in 
the observation data from \citet{Luri2020}. Even though these  
qualitative comparisons indicate the model reproduce well the observation proper motions, there is one difference 
between the model and data. For example, the proper motions of outskirts of the LMC 
have a bottom-right bulk motion in the model but a bottom-left pattern in the data.

\section{The Northern Tidal Arm}

In this section we examine the giant tidal arm associated with LMC, and its observational properties are reproduced by the \citet{Wang2019} model. 

\subsection{The general properties of observed and modeled Northern Tidal Arm}

\begin{figure*}
\includegraphics[scale=0.5]{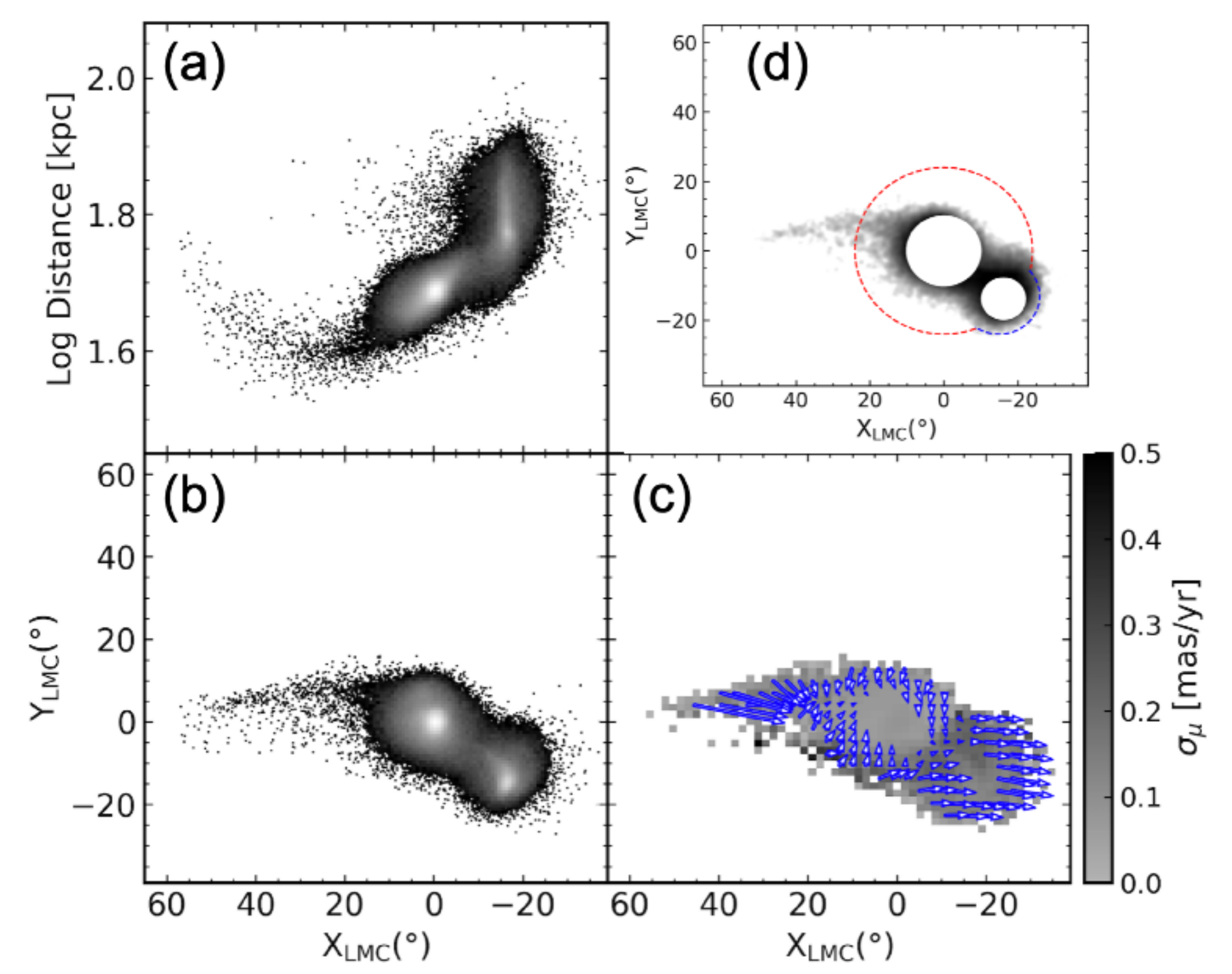}
\caption{The distribution of modeled MCs in the sky and distance maps. Panel a: particles distance distribution along X$_{\mathrm{LMC}}$ . 
 Panel b: particle distribution in the sky with orthographic projection \citep[see the definination in][]{Luri2020}. Panel c: particle distribution with color 
indicates proper motion dispersion comparing with $Gaia$ data \citep{Luri2020}, and blue arrows show the proper motion vectors. 
Panel d: number density distribution of simulation model. The red and blue dashed line indicate the selection region with Gaia 
EDR3 in \citet{Luri2020}.}
\label{fig:NTA_large}
\end{figure*}

By using the first year Dark Energy Survey data, \citet{Mackey2016} identified
the Northern Tidal Arm (NTA), which emanates from the edge of LMC, and
stretches more than 12.5 degree. This feature has been confirmed by further
studies \citep{Mackey2018,Belokurov2019}.  The stellar population of the NTA
matches well that of LMC, which is predominantly old with  $[Fe/H]$ $\sim$ -1.
This  indicates a LMC origin for the NTA.  $Gaia$ EDR3 provides deep
astrometric data, and in particular, proper motions may provide important
constraints about the NTA origin.

\citet{Luri2020} confirm this NTA feature and found that it stretches to more
than 20 degree, with NTA stars showing motions towards the LMC (see their
figure 17).  To compare with the $Gaia$ data, we project our simulation data
(see the panel b of Fig. \ref{fig:NTA_large}) to the same frame as
\citet{Luri2020}. Simulations predict a north giant stream starting from the
LMC edge, which match well with the NTA, though it stretches a larger angle,
$\sim$ 60 degrees. 

Since the discovery of NTA, several simulation models have been run to
explore the origin of NTA \citep{Mackey2016,Belokurov2019,Cullinane2022a}.  As
pointed out by \citet{Gatto2022} they only produced a more diffuse twisted
stream.  In the panel d of Fig.\ref{fig:NTA_large} we show the surface number
density distribution of our simulation model. Our model can produce a straight
stream that resembles the NTA. The NTA is thin in
particular at the tip, but at the edge of LMC the simulated NTA are thick,
which are likely associated to other substructures (see discussion in next
section).

Panel a of Fig. \ref{fig:NTA_large} shows the distance versus the
longitude distribution. The simulated NTA shows a gradient in distance, which
decreases from the edge of LMC to $\sim 38$ kpc at 30 degree away from the LMC,
and then increases to $\sim 56$ kpc at $\sim 55$ degree away from LMC.
It confirms that the NTA originates from the LMC. We note that the recent
observation from Magellanic Edges Survey \citep[MagES]{Cullinane2020,
Cullinane2022a} have confirmed an LMC origin for the NTA on the basis of its geometry,
metallicity and kinematics. 

In the simulation, the interaction between the Clouds triggers instabilities of
LMC disk, and then the disturbed LMC disk further suffers the strong Milky Way
gravity tidal field that forms a tidal arm and stretches it along when the LMC
falls into the MW.  This explains  well why NTA and LMC stellar population are
so similar.  The simulation thus predicts that there may be a much longer NTA
at lower surface brightnesses, which needs to be confirmed by deeper
observations.

\subsection{Comparing the morphology of Northern Tidal Arm and North-Eastern structure with observation}

\begin{figure*}
\center
\hspace*{-1cm}
\includegraphics[scale=0.37]{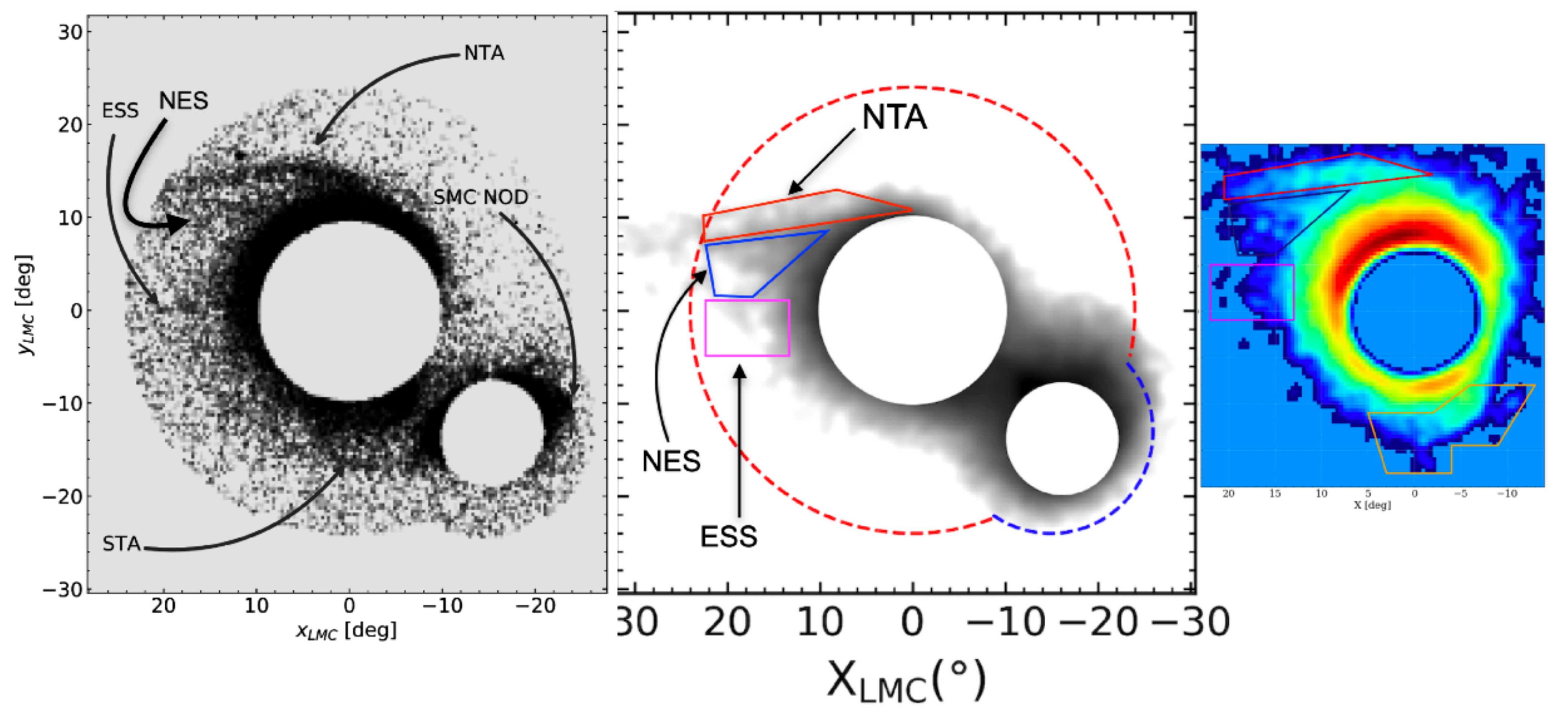}
\caption{Comparing the morphology of MCs with Gaia EDR3 data. The left
panel is from \citet{Luri2020}, the middle panel shows the \citet{Wang2019} model, and the right
panel is from \citet{Gatto2022} using Gaia EDR3 data with a Gaussian Mixture
Model to a sample of strictly selected candidate members of the Magellanic
System. In the middle panel, the red and blue dashed-line indicate the sample 
selection region for Gaia data \citep{Luri2020}.
In the right panel, the colored polygon regions indicate different 
substructures associated with LMC detected by \citet{Gatto2022}. 
The red, blue, and magenta polygon regions on right panel indicate the NTA, North-Eastern structure (NES), 
and Eastern Substructure (ESS), which are placed on the middle panel with corresponding  position.} 
\label{fig:NTA_morph} 
\end{figure*}

Fig.\ref{fig:NTA_morph} give a close comparison morphology of substructures
in the periphery of LMC.  The left panel shows the faint features  presented in
\citet{Luri2020} with $GAIA$ EDR3, in which the NTA is clearly identified as
well as several other substructures indicated by arrows, such as Eastern
Substructure \citep[ESS][]{ElYoussoufi2021}. Recenly, using a Gaussian Mixture Model to a strictly
selected sample of Magellanic System, \citet{Gatto2022} confirmed these
substructures, e.g., NTA (red polygon in the right panel of
Fig.\ref{fig:NTA_morph})  and ESS (magenta rectangle in the right panel of
Fig.\ref{fig:NTA_morph}). In the meanwhile,  they also find a new diffuse
sub-structure protruding from the outer LMC disc, which extends to more than 20
degree from the center of LMC.  The new feature is named North-Eastern
Structure (NES) and indicated by a blue polygon in the right panel of
Fig.\ref{fig:NTA_morph}. 

The middle panel of Fig.\ref{fig:NTA_morph} shows our model, which predicts both
the NTA and the NES. Besides these two structures, in the eastern part of LMC there is
substructure which could be associated with ESS. In the bottom left panel of
Fig.\ref{fig:NTA_large}, there are many particles along the NTA in the eastern
of LMC, which have the same origin as NTA and are induced by Galactic tides on the
LMC. Therefore, the \citet{Wang2019} reproduces NTA, NES, and  ESS. Future observations 
to measure the metallicity of NES and ESS are needed to compare with NTA and verify 
their origin.

\subsection{The distance to NTA}

\begin{figure}
\center
\hspace*{-1cm}
\includegraphics[scale=0.40]{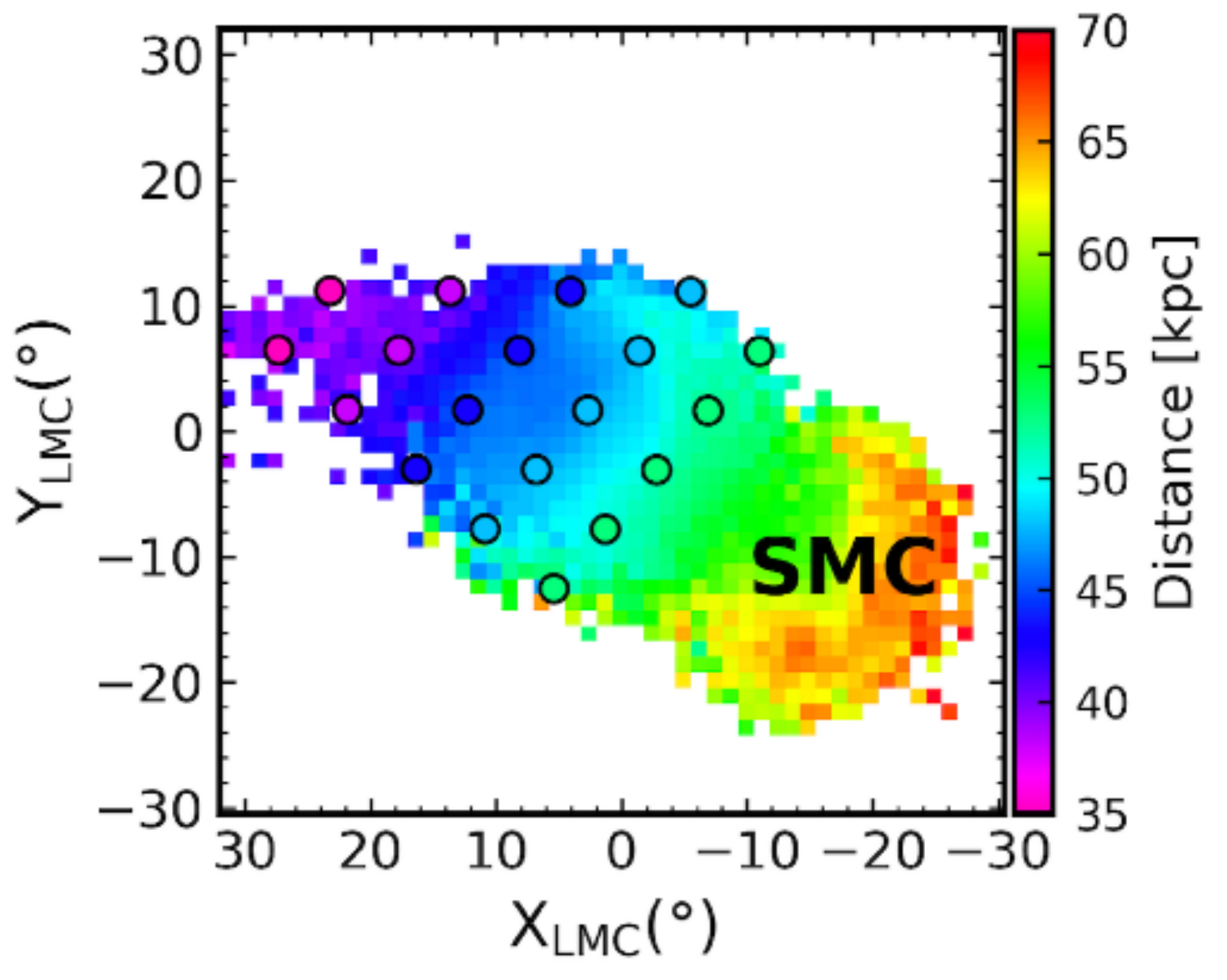}
\caption{The distribution of the line-of-sight distance for our modelled
MCs, and the colored circles show the distances for a disk with inclination
$i=34$ degree and orientation of 139.1 degree following \citet{vdM2014}. By examing 
the distances for a few observed fields associated with NTA with red clump stars, 
\citet{Cullinane2022a} found that the distance distribution of NTA follow the 
disk geometry of \citet{vdM2014} very well within $\sim 20$ degree. }
\label{fig:NTA_dis} 
\end{figure}

\citet{Cullinane2022a} using data from Magellanic Edges Survey (MagES)
\citep{Cullinane2020} and $Gaia$ EDR3 have studied the kinematic, metallicity
and distance for the NTA. Their fields cover to $\sim 20$ degree along NTA.
They found that the NTA is near the plane of the LMC disc, with
an inclination $i=34$ degree and an orientation $\Omega=139.1$ degrees
 \citep{vdM2014}. Fig.\ref{fig:NTA_dis} shows the line of sight
distance for our model, in which colored circles indicate an inclined disc
with a geometry following \citet{vdM2014}. Our modeled NTA follows well this
inclined disc within $\sim 20$ degree, which is consistent with measurements
by \citet{Cullinane2022a}.  

Fig.\ref{fig:NTA_dis} shows a gradient of distance in our
modeled SMC from the eastern to the western part of the SMC, with the western
part (see red color on the right) at more far distance than the eastern part (see green color on the left of the SMC).  In the observation,
there is indeed a distance gradient identified  with different tracers.
\citet{Muraveva2018} used 2997 RR Lyrae stars to study the three-dimensional
structure of SMC. They found that the line-of-sight depth of SMC is in the
range 1-10 kpc, and the eastern part of the SMC is located closer to us than
that of the western part.  \citet{Grady2021} used Gaia DR2 RR Lyrae to trace
the the three-dimensional distribution of MCs, and they arrived to the same
conclusion, i.e., the eastern portion of SMC lies at closer distance.
\citet{Scowcroft2016} used classical Cepheids to study the three-dimensional
structure of the SMC, and they found that the eastern side is up to 20 kpc closer
than its western side. The different distance gradient  depends on
the different stellar populations, which reflects the different morphologies and
line-of-sight distributions \citep{Ripepi2017}.  Even though detailed comparison
of the depth distribution for SMC between observation and our model will
require detailed stellar population modeling and sample selection, the Wang et al. model
predicts and reproduces the existence of this distance gradient from
eastern to western part of SMC.

\subsection{The kinematics of NTA}

\begin{figure*}
\hspace*{-1cm}
\includegraphics[scale=0.48]{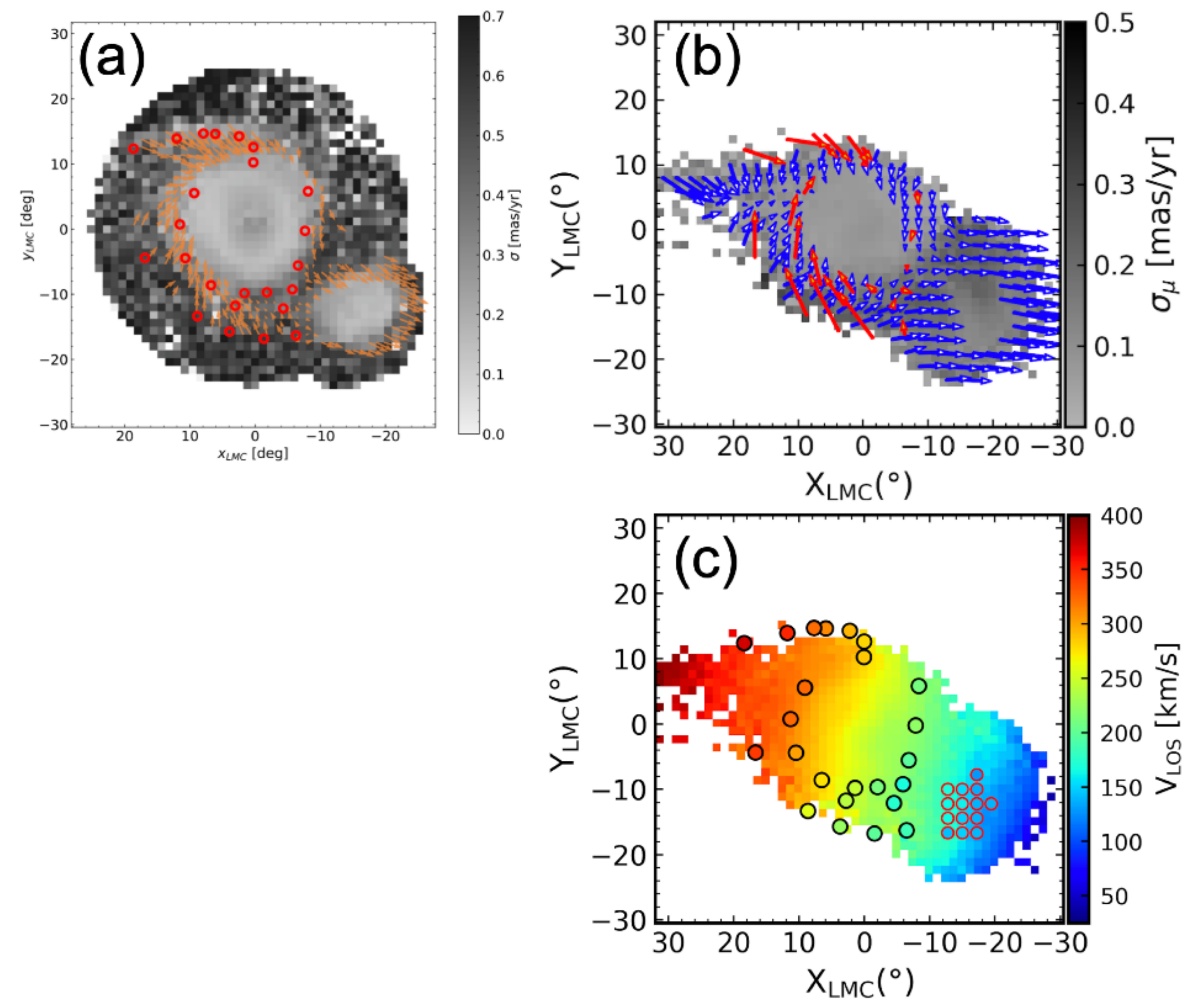}
\caption{The panel a is from \citet{Luri2020} with $Gaia$ EDR3 showing the
proper motion dispersion map and proper motion vectors shown by orange arrows.
We note that there is no proper motion vector scale length available from
\citet{Luri2020}. The red circles indicate the observed fields by MagES
\citep{Cullinane2020}. The panel b shows the simulated MCs with blue arrows indicating the modeled proper motions, and the red arrows indicate the proper motions derived from
MagES in each observed field. The panel c shows our modeled line-of-sight
velocity field, and big colored circles around LMC indicate the line-of-sight 
velocity from MagES \citep{Cullinane2022a,Cullinane2022b}. The small colored circles 
in the center of the SMC are observed line-of-sight velocities from literature data (see the text for details).}
\label{fig:NTA_kine}
\end{figure*}

Fig.\ref{fig:NTA_kine} compares the proper motions from Gaia EDR3 data 
\citep{Luri2020} (orange arrows in the panel a) with our model (panel b). 
The proper motions of stars in the simulated NTA (panel b of
Fig.\ref{fig:NTA_kine}) imply that they are moving towards the LMC, which matches well with
$Gaia$ results, except at the location closest of the LMC, for which model deviates
from the observations.  As in the observations, the dispersion of the proper
motions in the simulated NTA are small.   The length scale
of proper motion vectors are different between model and observations data because of the lack of
information about the amplitude of the observed proper motion vector \citep{Luri2020}.

With MagES \citep{Cullinane2020} combined with Gaia EDR3 data,
\citet{Cullinane2022a,Cullinane2022b} have selected members of Magellanic
Cloulds with fitting a multi-dimensional Gaussian distribution to the LOS
velocity and proper motion.  There are 26 selected fields observed in the
periphery of LMC as indicated by red circles in the panel a of
Fig.\ref{fig:NTA_kine}. With these data, they have derived the proper motions
for each field, which are shown by red arrows in the panel b of
Fig.\ref{fig:NTA_kine} for comparison with our modeled proper motion. The
proper motion of simulated NTA show downwards motion close to LMC disk, while
the observation data show bottom-right toward motion.  We note that the four
red arrows in the outmost bottom-left disk have very large vector length. This
could be due either to a small number of stars available for measurements from
MagES, or because they are associated with faint
substructures\citep{Cullinane2022b,ElYoussoufi2021} which are not reproduced by
current models.

With MagES, the LOS velocity for each field is also derived from the members
stars, which are shown by colored circles in the panel c of
Fig.\ref{fig:NTA_kine} for comparison with our modeled line-of-sight velocity
map.  The simulated LOS velocity field follows well that observed (colored circles).

We note that there is a radial velocity gradient across SMC with increasing radial
velocity toward LMC.  With new spectroscopic data of $\sim 3000$ RGB and
complemented by literature spectroscopic measurement, \citet{DeLeo2020} have
identified a large scale radial velocity gradient for SMC with increasing
velocity toward the Magellanic Bridge (see their Figure 7). To have a better
comparison with observations, we have collected observation data of radial velocities
in the SMC from literature, including $\sim 3000$ RGB stars from \citet{Dobbie2014}, and $\sim
2000$ from \citet{Evans2008} with majority of OBA type stars, as well as data
from \citet{DeLeo2020}.  These observed radial velocities in the SMC are shown
within the small colored circles in the panel c of Fig.\ref{fig:NTA_kine}. Even
though these observed data are focused on the central SMC, the observed
velocity gradient matched well with modeled one, confirming that the \citet{Wang2019} velocity field in
SMC follows the observational data.

\section{DISCUSSION}

Even though the MS has been discovered for about fifty years, there is still no agreement about its origin, which from the literature could be either a gigantic tidal tail, or two ram pressure tails trailing behind each of the two Clouds. Both origins are consistent with the fact that the MS is trailing behind the Clouds \citep{Mathewson1974}, and that the recent collision between the Clouds had formed the Bridge. However, the tantalizing amount of high precision data provided by $Gaia$ and ground-based surveys that is now available, is probably sufficient to disentangle which mechanisms lead to the MS formation. \\

In the following, we do not discuss the LA formation, which can be either attributed to the Cloud interaction \citep{Besla2012,Lucchini2020}, or to dwarfs passing ahead of the Clouds, which have lost their gas at different locations  explaining then the LA four arm behavior \citep{Hammer2015,Wang2019,Tepper-Garcia2019,Lucchini2021}. First, this is because forcing the LA to be of the same origin than the MS could lead to misleading results. Second, there are more evidences that the LMC is associated to dwarfs \citep{Patel2020}, and also that many MW dwarfs have a recent infall like the LMC \citep{Hammer2021}, which let plausible a formation of the four leading arms by passages of several leading gas-rich dwarfs.  \\

Table~\ref{tab:models} provides a list of the Magellanic System main properties, and describes the ability  of four modeling to reproduce them. We have chosen to describe only models based on a first passage of the Clouds, which let us with the tidal models of \citet{Besla2012}, \citet{Lucchini2020}, and \citet{Lucchini2021}, and with the 'ram-pressure plus collision' model of \citet{Wang2019}. Here we do not compare the model of \citet{Hammer2015}, because the \citet{Wang2019} model is its direct adaptation by just changing the $GADGET2$ by the $GIZMO$ software, for a better accounting of the Kelvin-Helmholtz (KH) instabilities. Such instabilities have to be accounted to constraint the MW halo gas content from the MS actual length and distance, as well as to properly account for the ionized gas kinematically associated to the MS \citep{Fox2014}.  \\

Ionized gas properties cannot be simulated by the \citet{Besla2012} model, since it does not include the MW halo gas, which cannot then interact with the predicted gaseous tidal tail. Because the \citet{Besla2012} model predicts a 10 times smaller HI mass than that observed in the MS is a major problem for this modeling, since adding the MW hot gas corona can only decrease the HI mass due to KH instabilities. However, this problem has been circumvented by \citet{Lucchini2020}, who further added a hot corona to the LMC, in order to account for the large ionized gas mass associated to the MS. \\

In Table~\ref{tab:models} we do not account for the mass properties of the remnant Clouds\footnote{Notice that the gas mass attached to the Clouds as well as the SMC shape are remarkably reproduced by \citet{Wang2019}}, because these properties can be tuned after modifying their initial conditions. However, the goal is to reproduce the whole MS properties, together with the Cloud orbital motions, for which measurements from $Gaia$ accurately define their orbits and the way they have deposited their gas to form the MS.  \\

Table~\ref{tab:models} also shows that the \citet{Wang2019} model reproduces all the main properties of the MS, while other modeling fail for at least half of them. However, as argued by \citet{Tepper-Garcia2019}, one may consider that the initial conditions rely on a large number of model-dependent, and thus necessarily tuneable parameters leading to costly numerical experiments, and then prohibitive (or even 'futile') to explore in full the available parameter space.  Following these considerations, one would only conclude that the 'ram-pressure plus collision' model is just the more advanced one by providing an explanation of all MS properties, but that a tidal model could in principle reach the same success.\\

However, this paper shows that the 'ram-pressure plus collision' model does not only reproduce the MS properties, but is also predictive, e.g., for the complex properties of stars in the Bridge, or for the NTA formation. Moreover, one may try to identify properties that cannot be reproduced by the tidal model. To do so, we propose to reinvestigate the physical properties of the MS as reported by \citet{Hammer2015} after their examination of the exquisite data from the Galactic All-Sky Survey \citep[GASS;][]{McClure-Griffiths2009,Kalberla2010}.  It revealed two inter-twisted filaments along the MS length, characterized by a transonic flow (see also \citealt{Bland-Hawthorn2007}) in a moderate to low turbulent medium (Reynolds parameter, $R_{e}$, of few hundred) . These last properties are consistent with the presence of vortices anchored into the inter-twisted filaments \citep{Hammer2015}. \\

Perhaps the major drawback of the tidal model is that it cannot reproduce the MS morphology (e.g., the inter-twisted filament behavior), including the fact that  kinematic and chemical analyses indicate that gas from both the LMC and SMC is present in the MS, as it is argued by \citet{Lucchini2020}. The \citet{Lucchini2020} simulations indeed include an additional filament that seems attached to the LMC as it is observed \citep{Nidever2008}, but the simulated MS HI morphology is so wide (see their Figure 2) that it show very few similarities with the observed narrow inter-twisted filaments \citep[see their Figure 2]{Hammer2015}. The major addition made by \citet{Lucchini2020} to the \citet{Besla2012} model from which it is adapted, is indeed to add a Galactic corona to the MW as well as another corona to the LMC. The advantage of the later addition is to reproduce by construction the large amounts of ionized gas following the MS. However, this is at the cost of predicting a larger mass for the hot corona associated to the LMC than that linked to the MW (3 $\times10^{9}$$M_{\odot}$  versus  2 $\times10^{9}$$M_{\odot}$, respectively), which appears unrealistic. \\

\begin{figure*}
\center
\hspace*{-1cm}
\includegraphics[scale=0.74]{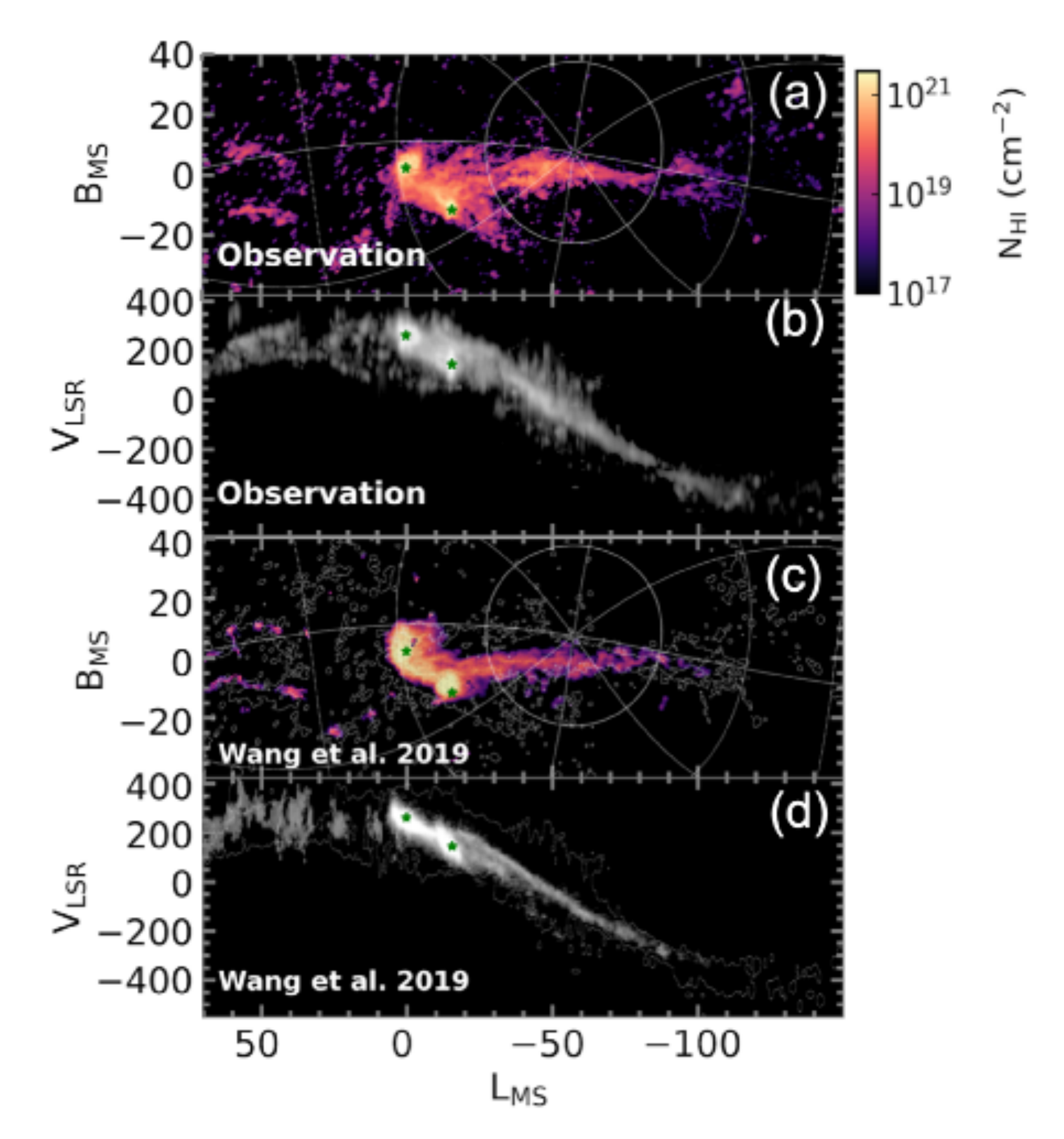}
\caption{Comparison of MS in the distribution of Magellanic longitude versus
latitude coordinates and longitude versus line-of-sight velocity. Panel (a) and
(b) show the observation results from \citet{Nidever2010}.  The panel (c) and
(d) show model results from the 'ram-pressure plus collision' model
\citep{Wang2019}, assuming Leading Arm formation from \citet{Hammer2015} with a 
scenario of gas deposited from several former leading gas-rich dwarfs. 
The observations and model can be compared with the panel (d) and (e) of Figure 3 of \citet{Lucchini2021}, 
and the panel (b) and (c) of Figure 2 of \citet{Lucchini2020} from tidal model with massive LMC corona. } 
\label{fig:diff_models} 
\end{figure*}

 Such a major difficulty has been identified and corrected by \citet{Lucchini2021}, but at the cost of changing dramatically the Cloud orbits, in such a way that their accurately observed velocities are not reproduced at $\ge$ 3$\sigma$ (and from 7 to 18 $\sigma$) for the LMC (SMC) tangential velocity components\footnote{In the notes to their Table 1, \citet{Lucchini2021} quoted that due to numerical resolution they found extremely large uncertainties (up to 0.5 mas yr$^{-1}$) in their simulated proper motions, a problem that surprisingly seems to occur only in the tangential direction, and which we never encountered in \citet{Hammer2015} or in \citet{Wang2019} simulations.}, respectively. Figure~\ref{fig:diff_models} shows the comparison of the 'ram-pressure plus collision' model (panel c and d) with observational data (panel a and b), the model is from \citet{Wang2019}. In panels c and d, the Leading Arm data from \citet{Hammer2015} have been added assuming the scenario of gas deposited from former runners of several gas-rich dwarfs. The model and observations can be compared with Figure 2 in \citet{Lucchini2020} and Figure 3 in \citet{Lucchini2021}.

 In summary, the 'ram-pressure plus collision' model succeeds to reproduce all the MS properties and is found to be also predictive, while the tidal model appears unable by construction to reproduce the very well defined inter-twisted filaments that constitute the MS. Moreover, tidal models that may reproduce the associated large amounts of ionized gas have led to inconsistencies either on the relative mass of the LMC vs MW coronas, or on the LMC and SMC tangential velocities. \\

\begin{table*}
\begin{center}
\caption{Comparing three modelings for reproducing the main properties of the Magellanic System}
\begin{tabular}{lcccc}
\hline \hline
                                    & Tidal Model       & Ram-pressure+Collision & Tidal+LMC corona     &  Tidal+LMC corona  \\
Observation Constraints             &\citet{Besla2012}  & \citet{Hammer2015}     & \citet{Lucchini2020} &  \citet{Lucchini2021}   \\
		                    &                   & \citet{Wang2019}       &                      &                        \\
\hline
MS ionized gas mass                 & N                & Y                       & Y                 &  N                    \\
MS ionized gas sky distribution     & N                & Y                       & Y                 &  Y                    \\
MS HI total mass		    & N	               & Y                       & Y       &  Y                    \\
MS HI gas velocity distribution & Y               & Y                       & N                &  Y                    \\
two inter-twisting filaments 			    & N                & Y                       & N         &  N            \\
No stars in the MS                  & N                & Y                       & N         &  N                   \\
\hline                                                                                              
Cloud proper motions                   & Y                & Y                       &  Y                &   N                   \\
\hline                                                                                                             
HI bridge                           & Y                & Y                       &  Y                &   Y                   \\
Young and old stellar bridge        & Y                & Y                       &  ?         &   ?          \\
\hline                                                                                                             
\hline
\end{tabular}
\end{center}
\label{tab:models}
\end{table*}

Table~\ref{tab:models} points out the fact that the ''ram-pressure plus collision'' model naturally reproduce as much as possible the numerous properties of the Magellanic System. This could be considered as natural, since the LMC HI disk has been undoubtedly shrunk by ram pressure effects \citep{Nidever2014}, and such a gas is expected to be trailing to form the LMC contribution to the MS. The predictive ability of this model (MB population and NTA) further indicates that this model goes into the right direction to disentangle the mystery of Magellanic System formation (Don Mathewson, 2015, private communication). The only possible 'caveat' of the ''ram-pressure plus collision'' model is that it predicts a moderate LMC mass to let the HI gas being extracted by ram-pressure to form the neutral Stream, especially between the Bridge and the tip-end.  \citet{Wang2019} simulations are based on small masses for the LMC, and they mentioned having failed in reproducing the MS for LMC masses larger than 2 $\times10^{10}$ $M_{\odot}$. In fact, the more massive is the ram-pressurized galaxy, the more difficult is it to extract neutral HI gas \citep{Yang2022}. However, a full numerical study is needed to estimate the highest  mass for the LMC halo that would not prevent the formation of the HI Magellanic Stream.



\section{CONCLUSION}

Here we show that the 'ram-pressure plus collision' model \citep{Hammer2015,Wang2019} is able to reproduce all the MS properties, as well as to predict new features that have been observed after the model elaboration, without fine tuning. The new observation features include the complexity of the stellar populations in the MB and the NTA. The MB is likely caused by the Cloud collision 200-300 Myr ago, for which the LMC has tidally extracted large amounts of gas from the SMC, the MB gas being then affected by the ram pressure exerted by the MW corona. \citet{Wang2019} explained as such the spatial and kinematic behavior of both Young Main Sequence stars and ancient RR Lyrae stars \citep{Belokurov2017}. 
Besides this, the stellar populations found at two different distances in the MB region by \citet{Omkumar2020} are also predicted by the \citet{Wang2019} model. The model also predicts that the foreground population results from the Cloud tidal interaction (mostly stars extracted from the SMC), while the background population is associated to SMC spheroid stars that have been tidally extracted and reshaped by the LMC. 
We have also compared proper motions from $Gaia$ EDR3 \citep{Luri2020} between young stars and the RC old stars in the MB region with \citet{Wang2019} model predictions. Both observations and model show that stars in the Bridge are moving consistently to the LMC. 
Recent identifications of the NTA and of its kinematics from $Gaia$ EDR3 have been also reproduced by \citet{Wang2019} model, which infers its origin from the LMC tidal stretching by MW tidal force. 

The ability of the 'ram-pressure plus collision' model contrasts with that of the tidal model that essentially fail to reproduce half of the main properties of the MS. In particular, the HI MS is mostly made of two inter-twisted filaments, which tidal models fail to reproduce, and for which no interpretation can be foreseen if the MS is a tidal tail. Moreover, the tidal model has difficulties to reproduce the MS gas mass, especially its dominant phase, the ionized gas, for which the proposed solutions appear either unrealistic, or with strong deviations from the calculated Cloud velocities. It is then likely that the MS is made by  ram pressure exerted by the MW corona to the Clouds since their entrance into the halo. This, combined with the 200-300 Myr collision that is robustly determined from the $Gaia$ proper motions of the Clouds, from their common star formation history, and from the Bridge, suffices to explain the whole MS properties. We further conjecture that to form the Magellanic Stream, the LMC mass has to be smaller than 2 $\times10^{10}$ $M_{\odot}$, though further studies are needed to precise the exact mass range.  



\section*{Acknowledgments}

We thank the referee for his/her helpful and insight comments, which help to
improve the manuscript.  The computing task was carried out on the HPC cluster
at China National Astronomical Data Center (NADC). NADC is a National Science
and Technology Innovation Base hosted at National Astronomical Observatories,
Chinese Academy of Sciences. This work is supported by Grant No. 12073047 of
the National Natural Science Foundation of China. We are grateful for the support of the International Research Program Tianguan, which is an agreement between the CNRS in France, NAOC, IHEP, and the Yunnan Univ. in China.

\section*{DATA AVAILABILITY}

The data underlying this article will be shared on reasonable request
to the corresponding author.

\bibliographystyle{mn2e}
\bibliography{reference.bib}

\begin{appendix}

\section{Gaussian fitting to the distance distribution}

\begin{figure*}
\center
\includegraphics[scale=0.45]{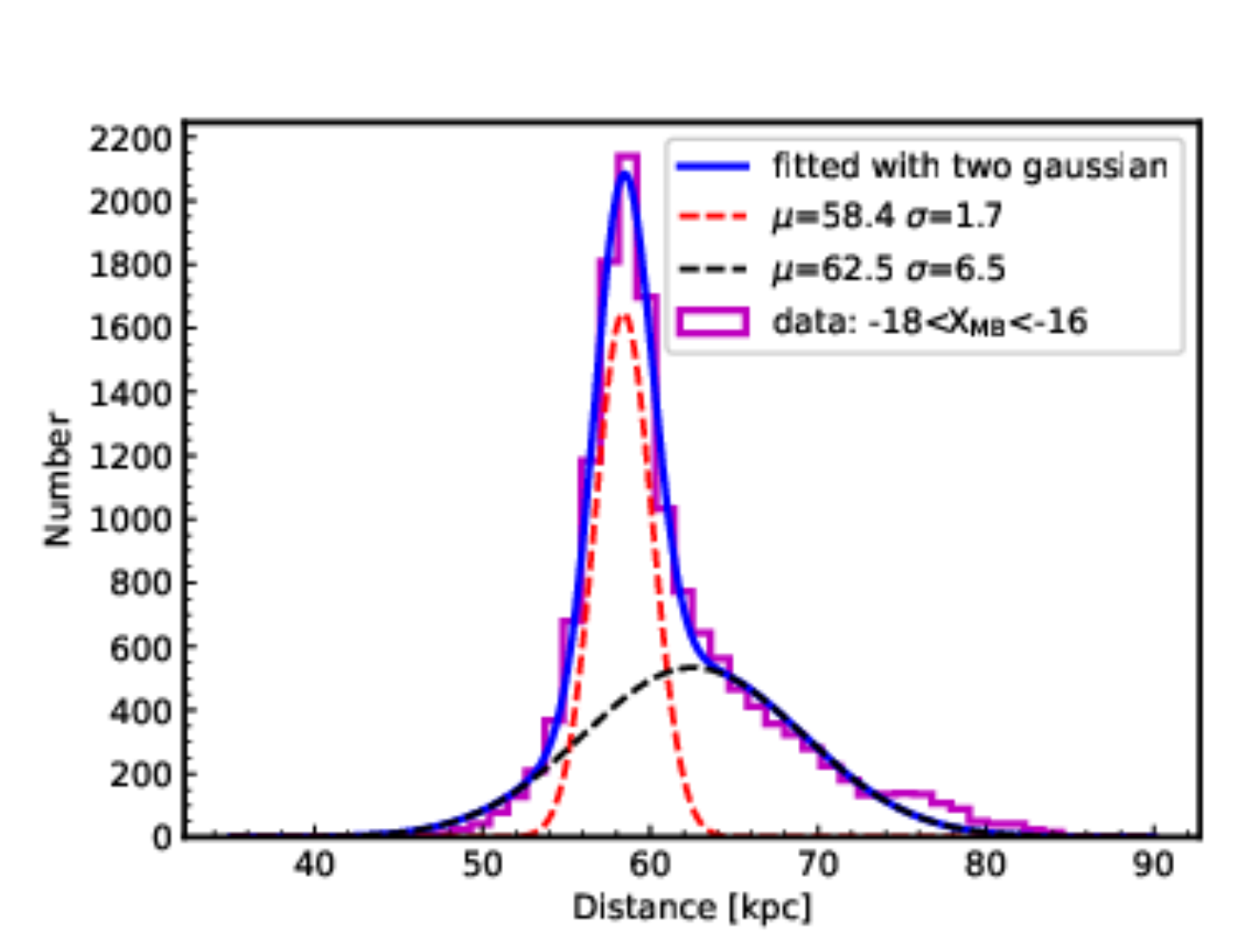}
\includegraphics[scale=0.45]{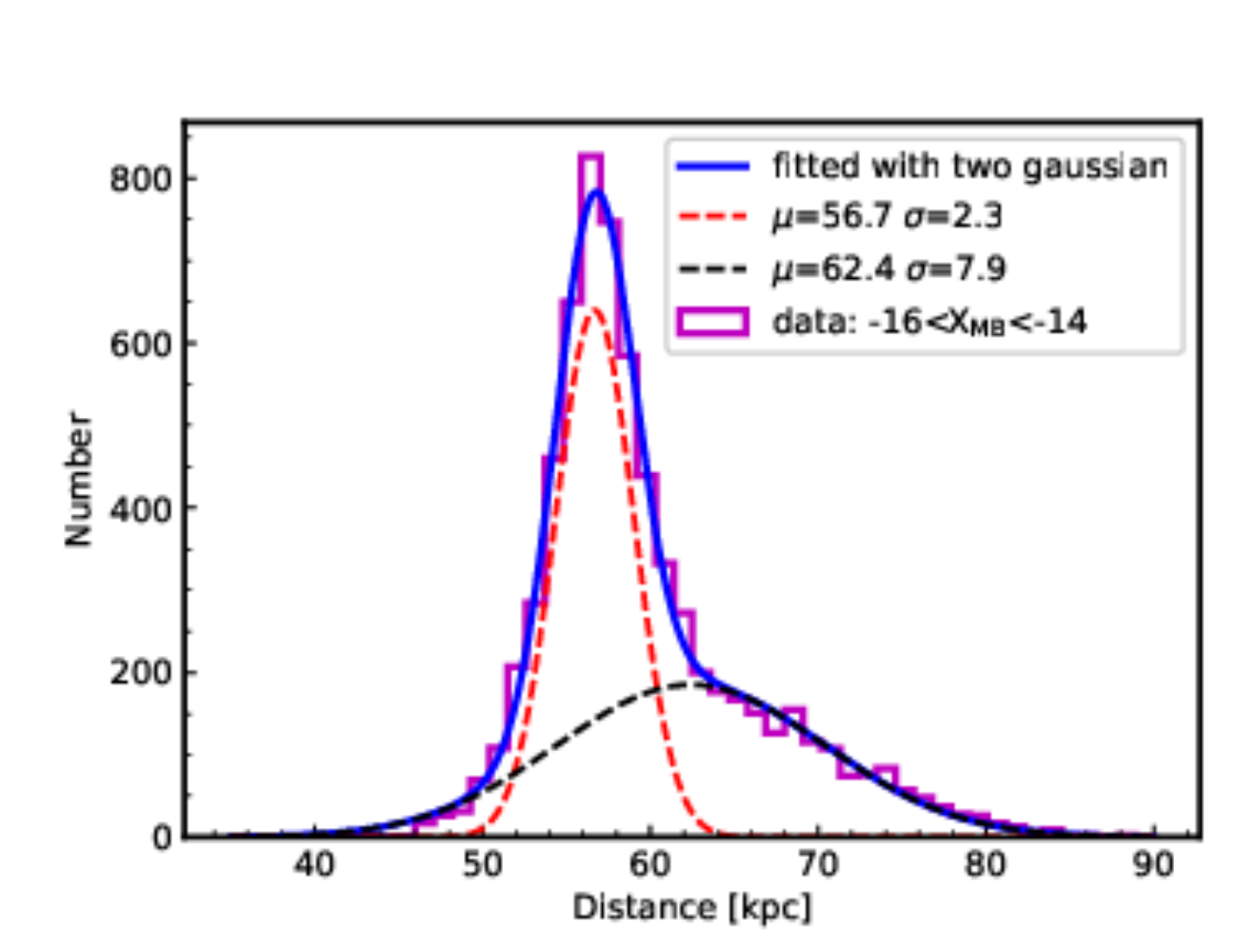}
\includegraphics[scale=0.45]{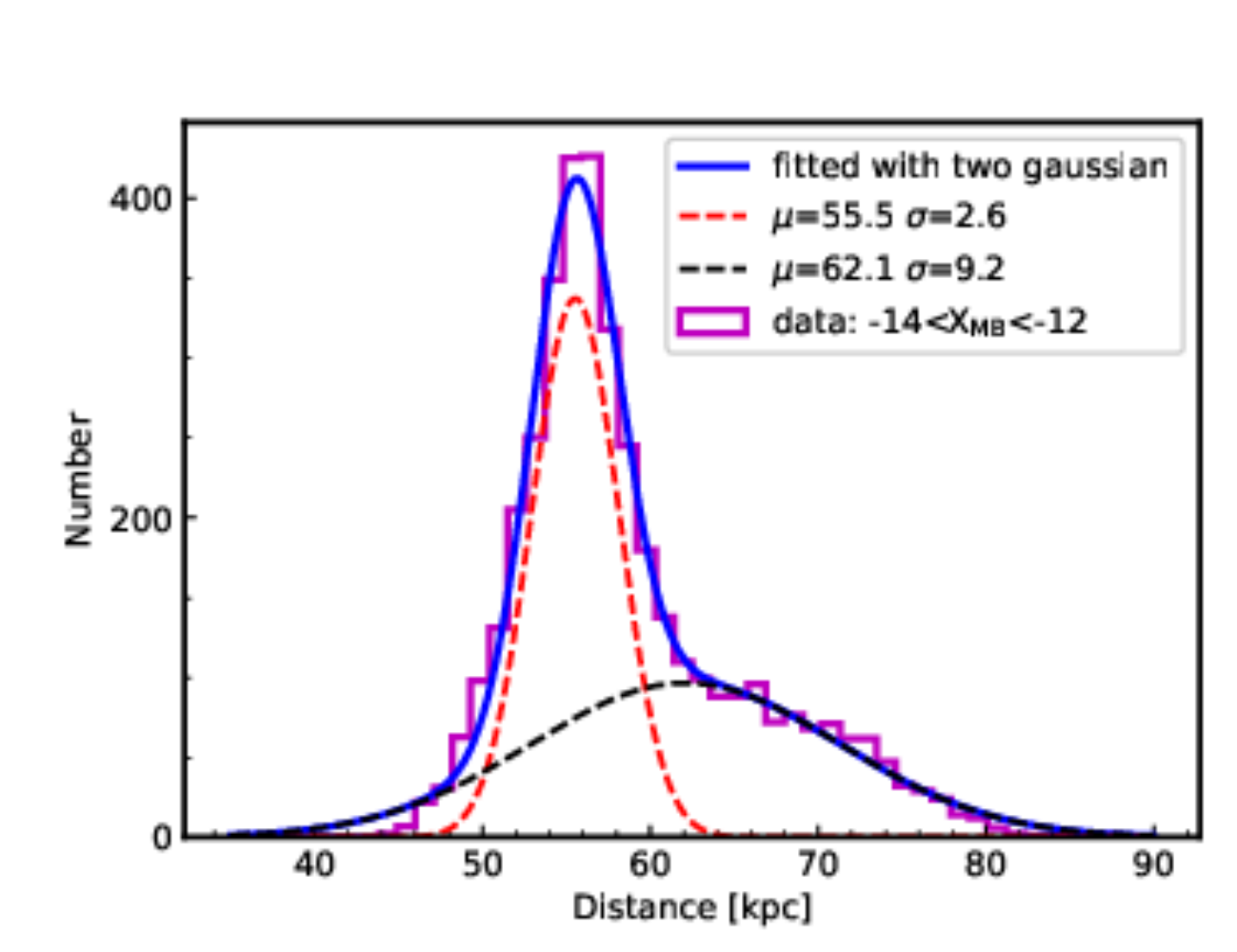}
\includegraphics[scale=0.45]{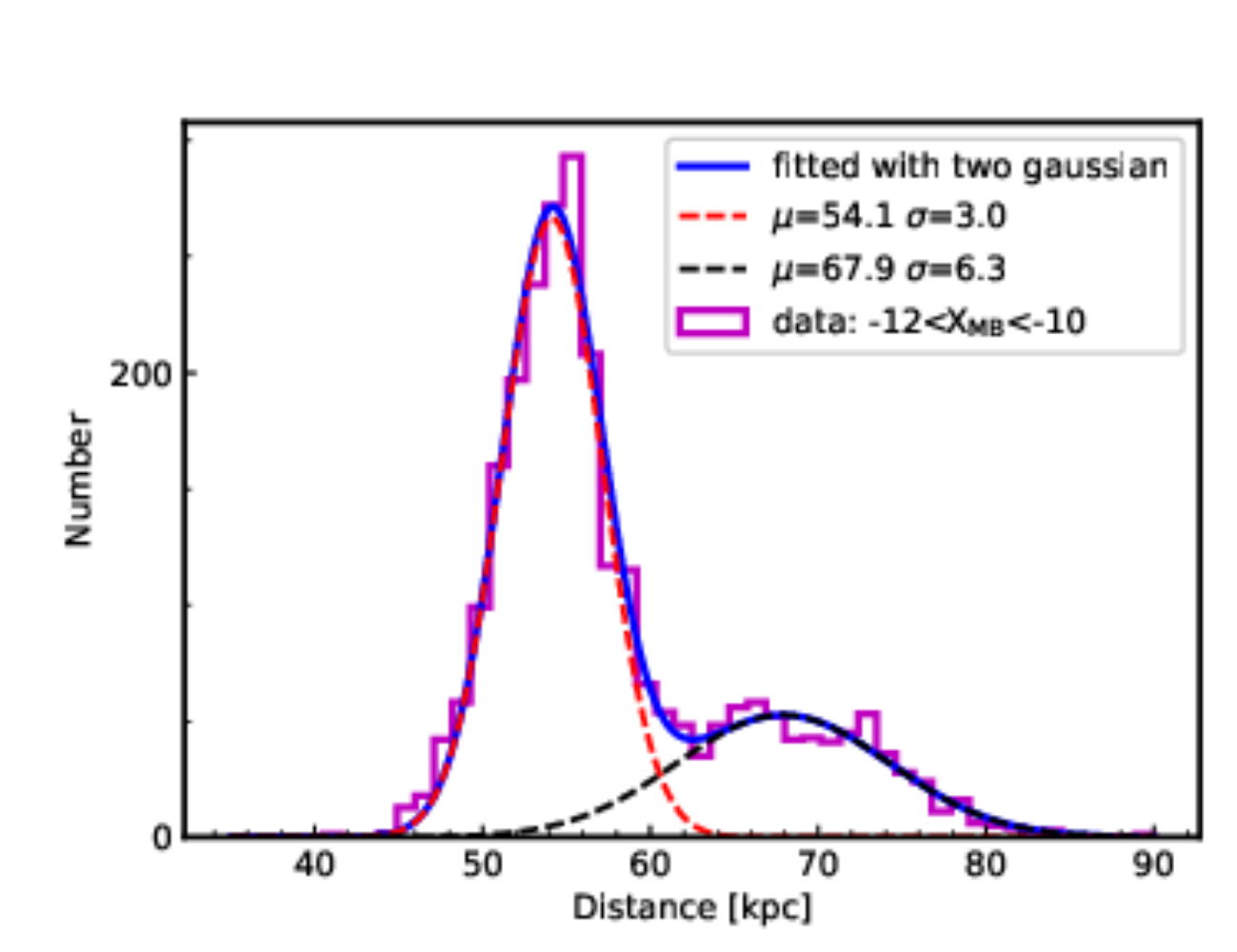}
\includegraphics[scale=0.45]{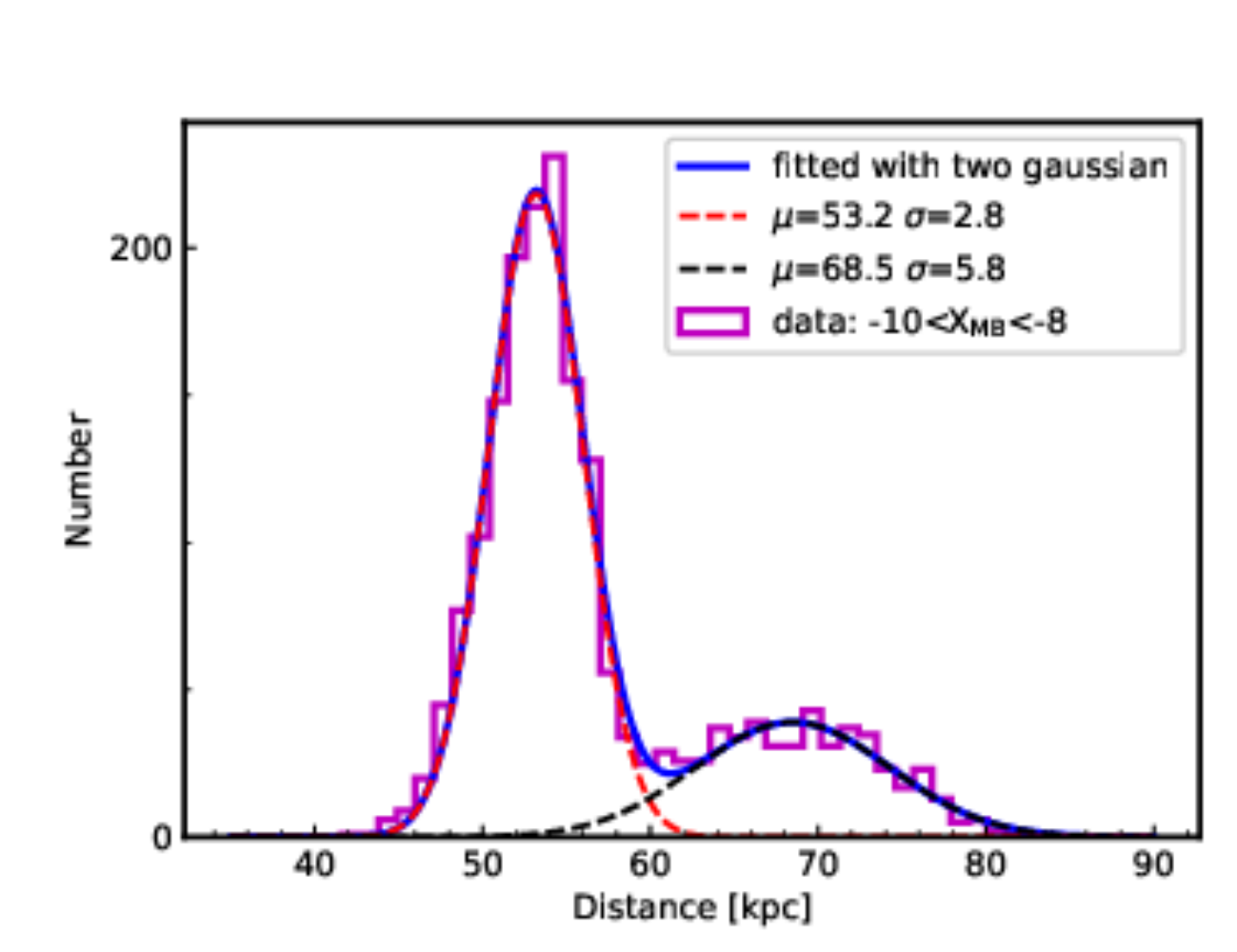}
\vspace*{4cm}
\caption{Distances distributions for particles (magenta) within different MB
longitudes intervals. Two Gaussian functions are used to fit the distance
distributions for foreground (red-dashed line) and background (black-dashed
line) population, with their mean and standard deviation are labeled on the top
right of each panel. The blue-solid line indicate the sum of two Gaussian functions.} 
\label{fig:hist}
\end{figure*}

Two Gaussian function are used to fit the distance distribution of particles 
within different MB longitude bins as shown in Figure \ref{fig:hist}.

\end{appendix}

\end{document}